\documentclass[nofootinbib,aps,twocolumn,showpacs,amsmath,amssymb,floatfix,superscriptaddress]{revtex4}
\usepackage{graphicx}
\usepackage{color}
\graphicspath{ {Figures/} } 

\begin{document}

\title{Spatio-temporal representation of long-delayed systems: an alternative approach}

\date{\today}

\author{Francesco~Marino}
\affiliation{CNR - Istituto Nazionale di Ottica, 
                largo E. Fermi 6, I-50125 Firenze, Italy}

\author{Giovanni~Giacomelli}
\affiliation{CNR - Istituto dei Sistemi Complessi, 
                via Madonna del Piano 10, I-50019 Sesto Fiorentino, Italy}

\begin{abstract}
Dynamical systems with long delay feedback can exhibit complicated temporal phenomena, which once re-organized in a two-dimensional space are reminiscent of spatio-temporal behavior. In this framework, normal forms description have been developed to reproduce the dynamics and the opportunity to treat the corresponding variables as true space and time has been since established. However, recently an alternative approach has been proposed in Ref. \cite{Marino2018} with a different interpretation of the variables involved, which takes better into account their physical character and allows for an easier determination of the normal forms. In this paper, we extend such idea and apply it to a number of paradigmatic examples, paving the way to a re-thinking of the concept of spatio-temporal representation of long-delayed systems. 
\end{abstract}

\maketitle

\section{Introduction}
A long-delayed dynamical system is characterized by a feedback action, which acts by re-injecting far-in-the-past information from the system itself. Notably, the time interval between the "present" and the "past" is assumed to be much longer than any other characteristic time-scale of the system without feedback ({\it long-delay limit}). Such a condition, apparently quite specific, appears naturally in disparate phenomena and the topic has attracted considerable attention in the last years (for a review, see e.g. \cite{Yanchuk2017}). Dynamical systems with long delayed feedback can display a rich variety of complex phenomena \cite{Chaos2017}. Such richness derives from the high dimensional phase space, and it is witnessed by scaling relations for extensive quantities analogous to those found in one-dimensional (1D) setups \cite{Farmer1982}. 

The standard approach to spatio-temporal modeling of long-delayed systems stems from the proposal introduced in Ref. \cite{Arecchi1992}. There, a two-dimensional (2D) coordinate system was used where a continuous variable $\sigma$ ranged in a delay interval played the role of a pseudo-space variable. The correspondent pseudo-time variable was the discrete index $\theta$, numbering the sequence of consecutive, disjoint delay intervals in the time series. This procedure, called \emph{spatio-temporal representation} (STR) amounts to express the time variable $t$ as
\begin{equation}
t = \sigma +\theta T~,
\label{STR_def}
\end{equation} 
where $T$ is the delay time. While this mapping is always feasible independently of the delay value, it is under specific circumstances that it shows its usefulness. One of the most important is that the system is actually operating in the long-delay limit: it is indeed in this case that the system evolves on two well-separated timescales, $\sigma$ and $\theta$, which thus effectively act as mutually independent variables \cite{Yanchuk2017}.

The above representation was quite successful and allowed to disclose many relevant features, common to the delayed and spatially-extended systems \cite{Giacomelli1996}. From the 1990s up to now, a number of experiments have been realized, in particular in the field of optics, demonstrating different kinds of equivalent spatiotemporal phenomena hidden in the temporal dynamics. These include e.g. defect propagation \cite{Giacomelli1994}, domain coarsening and nucleation \cite{Giacomelli2012,java2015,Giacomelli2013}, front pinning and localized structures\cite{Marino2014-2017,garbin2015,Romeira2016,Marinochaos}, chimera states \cite{Chimeras} and critical phase transitions \cite{Faggian}. Recently, even generalizations involving two, hierarchically long delays have been considered leading to the evidence of spiral defects and defects turbulence \cite{Yanchuk2014-2015}, 2D chimeras and dissipative solitons \cite{brunnerchaos} and excitable waves \cite{Marino2019}. The emergence of such a wealth of pattern structures confirms the role played by the multiple timescales in the long-delayed dynamics, supporting their natural identification as the main independent variables of the system.

However, recently a critical analysis of this approach has been reported \cite{Marino2018}, introducing an alternative representation for the data generated from long-delayed systems and suggesting a different spatiotemporal interpretation. In this framework the bulk dynamics is described in terms of a new rule,
the so-called {\it dynamical representation} (DR), employing the opposite definition of pseudo-space and -time variables with respect to the STR. The analysis in \cite{Marino2018} has suggested that, while the two representations are equally effective in evidencing pattern structures, a physical description in terms of a spatiotemporal model is more properly obtained in the DR.

In this work, we extend these results and compare the two representations in several respects, both on the basis of general arguments and with the help of a few of paradigmatic examples.

The plan of the paper is the following. In Sec.II we recall the main features of the representations remarking the conditions in which they are valid. In Sec. III, DR and STR are compared with respect to causality, first in terms of validity of Kramers-Kr\"onig relations, and then evaluating the comoving Lyapunov exponents. In the subsequent sections we analyze the two representations in the framework of two specific examples. In Sec.IV, we focus on the delayed Adler equation, describing the phase dynamics of an optically-injected laser system with feeedback \cite{garbin2015,yanchukPRL2019,julien2020}. In Sec. V, we will treat a model of passively mode-locked external-cavity surface-emitting laser recently introduced in \cite{schelte2019,schelte2020}. In Sec. VI, we will discuss show how to move between the representations on the basis of parity arguments. We will draw our conclusions and present some perspectives in the final section. 

\section{Definition of the representations}
In this Section, we start by recalling the concepts at the basis of STR and DR. Without any loss of generality we consider the scalar system 
\begin{equation}
u_t = F(u,u_d)~,
\label{delay}
\end{equation}
where $u_d(t) = u(t-T)$ is the delayed variable and $T$ is the delay time. The model (\ref{delay}) has to be accompanied with an initial condition specified on an interval of length $T$,  e.g. 
 \begin{equation}
u(t) = u_0(t)~,~ t \in [-T,0]~.
\label{delay_ic}
\end{equation}

As mentioned before, for a meaningful spatio-temporal representation of (\ref{delay}-\ref{delay_ic}) the delay time $T$ should be longer than any other timescale of (\ref{delay}) without delay. Such condition is necessary but not sufficient: we should also consider an observation time $t_{TOT} \gg T$ for an appropriate definition of the {\it thermodynamic limit}
\begin{eqnarray}
T                                   & \to & \infty \\
S  =  \big[ \frac{t_{TOT}}{T} \big] & \to & \infty ~,
\end{eqnarray}   

where $[.]$ stands for the integer part.

For the sake of clarity we introduce different sets of names for the variables involved in the two representations. We write (\ref{STR_def}) in the form
\begin{equation}
t = x +y T~,
\label{pattern_def}
\end{equation} 

where we refer to $x$ as the {\it fast time} and to $y$ as the {\it slow time} and we define the field $\Phi(x,y) = u(t)$. 

In the limit $T \to \infty$, the time derivative can be expressed as
\begin{equation}
\frac{d}{dt} = \partial_x + \frac{1}{T} \partial_y \to \partial_x ~.
\label{time_deriv}
\end{equation} 

This condition holds in the absence of the so-called anomalous Lyapunov exponent \cite{Giacomelli1995} or, equivalently in the weak-chaos regime \cite{Heiligenthal2011}, and amounts to state that the variations of $\Phi(x,y)$ along the $y$ direction are negligible asymptotically. In simple terms, $\Phi(x,y)$ should exhibits small variations between two successive delay units. Accordingly, the integer variable $y$ will be embedded into a real domain. Such an assert implies that there exist a correlation length $L_y$ of the pattern along the slow time such that $L_y > 1$. The field $\Phi(x,y)$ does not vary significantly on a scale $\Delta y=1$ along $y$, which results into a smooth pattern since several discrete points fall within a correlation length $L_y$.

Notably, the above requirements stay at the basis of both representations. They indicate whether the two timescales behave as mutually independent variables and thus can be used to parametrize a 2D smooth pattern. 

On the basis of these considerations, it is clear that the above re-organization of data in itself does not provide any constraint on the physical role of the variables in generating the dynamics.
These are actually introduced in the framework of the two representations, where the original model (\ref{delay}) is re-written in terms of the new variables $x$ and $y$, in order to build a suitable two-dimensional rule. 

Setting $x=\sigma$ and $y=\theta$, and defining $U(\sigma,\theta)=u(t)$, the model (\ref{delay}) reads 
\begin{equation}
\partial_\sigma U = F(U,U(\sigma,\theta-1)),~~\sigma \in [0,T]~,
\label{STR}
\end{equation}

together with the boundary conditions
\begin{eqnarray}
U(\sigma,-1) &=& u_0(\sigma)~,~ \sigma \in [0,T]  \nonumber \\
U(\sigma+T,\theta) &=& U(\sigma,\theta+1)~.
\label{STR_bc}
\end{eqnarray}

Eqs. (\ref{STR})-(\ref{STR_bc}) correspond to the standard spatio-temporal description of the delay model in the STR: the variable $\sigma$ is interpreted as the pseudo-space and $\theta$ as the pseudo-time. In particular, the smoothness of the pattern along $\theta$ allows to approximate $U(\sigma,\theta+1) \approx U(\sigma,\theta)$, leading to $U(\sigma+T,\theta) \approx U(\sigma,\theta)$, similarly to the periodic boundary conditions for a 1D spatially extended system.

In order to provide an effective mapping of the delayed dynamics, the next step is to employ Eq. (\ref{STR}) to derive an explicit rule for the pseudo-time evolution (i.e. along the $\theta$ direction). This can be achieved by means of different methods. As we have seen, the pseudo-spatial and pseudo-temporal variables are related the multiple timescales of the system, the fast time and slow time. A multiscale approach separating such scales into different perturbation orders is often very convenient, and allows to derive a partial-differential equation (PDE) able to reproduce the delayed dynamics in the $(\sigma,\theta)$ domain, obviously within some degree of approximation \cite{Giacomelli1996,Kashchenko1998,Yanchuk2014-2015,Wolfrum2006,Giacomelli1998,Bestehorn2000}.

The DR is an alternative approach to the STR, proposed in \cite{Marino2018}. It considers the opposite dynamical role for the two variables. In this scheme, we name $x=\tau$ as the pseudo-time and $y=\xi$ as the pseudo-space, defining a new field variable $Z(\xi,\tau) = u(t)$. The evolution rule derived from Eq.(\ref{delay}) is now written as
\begin{equation}
\partial_\tau Z = F(Z,Z_{NL})~,
\label{DR}
\end{equation}
where the delayed term translates into the {\it non-local} asymmetric spatial coupling $Z_{NL}(\xi,\tau) = Z(\xi -1,\tau)$ and the temporal evolution is along the former pseudo-space. 
Eq.(\ref{DR}) should also be complemented with suitable boundary conditions. Here we consider spatially-periodic boundaries conditions 
\begin{eqnarray}
Z(\xi,0) &=& z_0(\xi)~,~ \xi \in [0,S]  \nonumber \\
Z(\xi + S,\tau) &=& Z(\xi,\tau)~,
\label{DR_bc}
\end{eqnarray}

in the thermodynamic limit. We remark that from a strict mathematical point of view the correct solution of the original delay problem would be obtained only for one choice of the initial and boundary conditions (generally different from the periodic ones here used).  
We expect however that in the thermodynamic limit even an arbitary choice would produce patterns well approximating the delayed dynamics. In particular, we will see that this is indeed the case whenever conditions (\ref{DR_bc}) hold.

\begin{figure}
\begin{center}
\includegraphics*[width=1.\columnwidth]{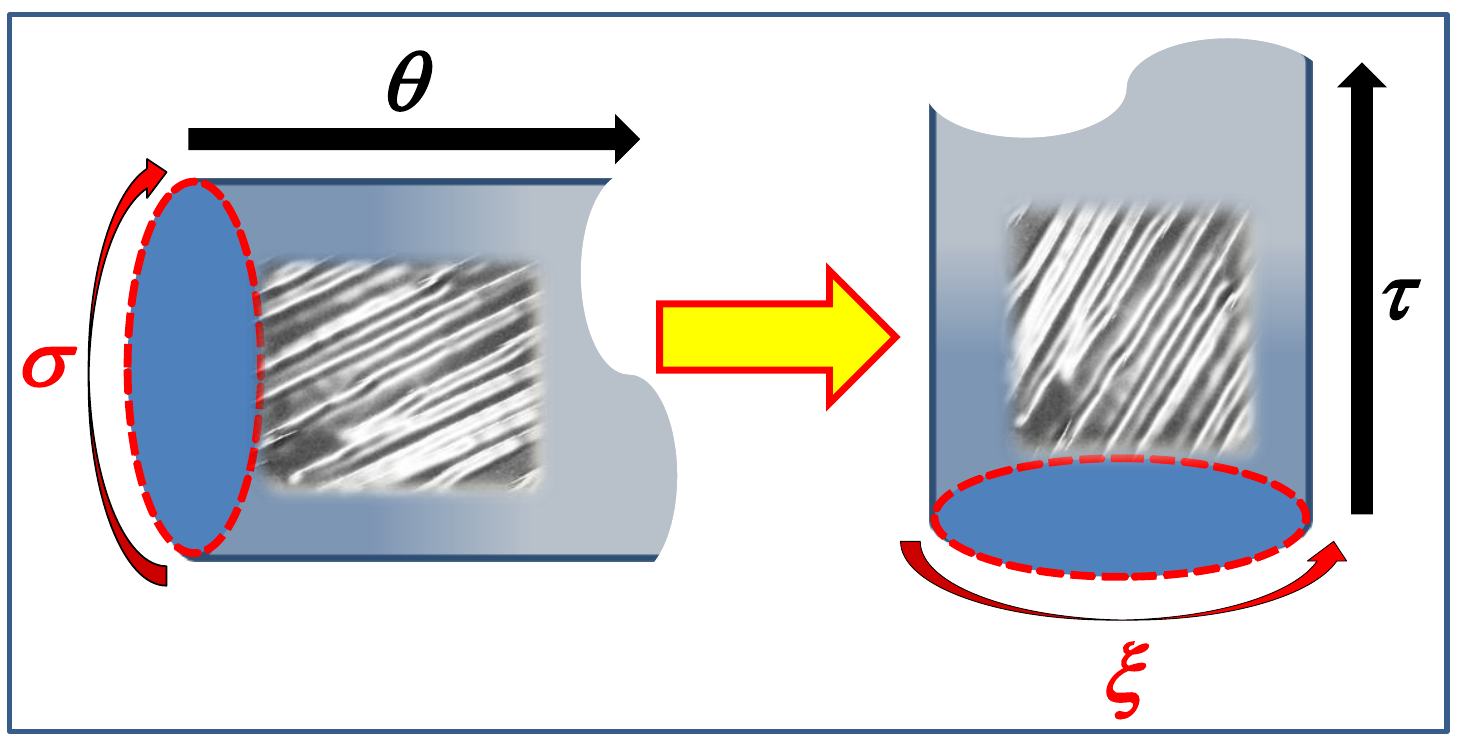}
\end{center} 
\caption{Pictorial view of the long delay pattern embedded into the STR (left) and DR (right) manifolds.
The dashed circular lines mark the initial conditions. The curved and straight arrows indicate respectively the periodic boundary conditions and the direction of evolution.}
\label{2appr}
\end{figure}

The topology of the variable domains associated to the two representations is illustrated in Fig. \ref{2appr}, evidencing different global manifolds. The dashed circular lines mark the initial conditions, the cylinder axis defines the direction of evolution (pseudo-time axis) and the cross-sectional circumference corresponds to the size of the spatial cell. The patterns produced in either one of the two representions can be readily identified looking at the location of the initial conditions and/or spatial boundaries. On the other hand, in the {\it bulk} region we will observe essentially the same dynamics since the rule generating the pattern far from space-time boundaries remains the same. 

Interestingly, the representation (\ref{DR}) also allows for a straightforward expansion of the non-local coupling in terms of spatial derivatives, leading eventually to a normal form description through standard PDEs
\begin{equation}
Z(\xi-1,\tau) \approx Z(\xi,\tau) -Z_\xi(\xi,\tau) +\frac{1}{2}Z_{\xi\xi}(\xi,\tau)-..~,
\label{expansionDR}
\end{equation}

where $Z_\xi = \partial_\xi Z,~ Z_{\xi\xi} = \partial^2_{\xi\xi} Z, ..$, obtaining the PDE
\begin{equation}
Z_\tau = \mathbf{F}(Z,Z_\xi,Z_{\xi\xi},..)~.
\label{pde}
\end{equation} 

As discussed before, the validity of the formal expansion (\ref{expansionDR}) relies on the assumption that the pattern exhibits small variations along $\xi$, i.e. that the correlation along $\xi$ decays over a length $L_\xi \gg 1$. In this case, a PDE model (PDEM) where the time derivative is explicitly written in terms of the spatial derivatives can be obtained directly, expanding the non-local term up to a given order and approximating the delayed dynamics with arbitrary precision. This represents an advantage with respect to the STR, in which the derivation of a PDE model often requires long calculations and the vicinity to a bifurcation.
In the specific case of a linear delay term, each order of the expansion can be associated to a specific physical effect: the zero-order is a renormalization of the local force, the first provides the advection (that can be removed with a suitable choice of a comoving reference frame), the second is diffusion, the third corresponds to dispersion, etc.

We conclude this section remarking that any different choice of the reference frame in the plane $(x,y)$ can be chosen to accordingly rewrite the {\it bulk} rule (\ref{delay}) and reproduce the pattern $\Phi$. Such models will be mathematically equivalent, all sharing a {\it non-local} coupling arising from the delayed feedback and a fairly good independence on the boundary conditions in their correspondent thermodynamic limit. However, they might give rise to physical inconsistencies. In the next section we will compare the STR and DR with respect to causality.

\section{Representations and causality}

The core point at the basis of a dynamically correct, spatio-temporal representation is whether the resulting mathematical model not only is capable to produce the embedding pattern $\Phi(x,y)$, but its variables can play the role of well-behaving space and time coordinates. While no special constraints can be assumed on the spatial variable, the temporal one must satisfy causality, i.e. the evolution along it must depend on its previous values only (the {\it past}). Our aim here is to investigate the causal structure of the two representations and thus of their associated spatiotemporal PDE models. 

\subsection{Susceptibility}

For a general linear system the notion of causality is equivalent to satisfy Kramers-Kr\"onig relations relating the real and imaginary parts of the complex susceptibility function \cite{toll}. 
We thus consider the linear long-delayed equation
\begin{equation}
\dot{X} = AX +BX_d~,
\label{lindel}
\end{equation}
where $X$ is a vectorial variable, $A$ and $B$ the matrices of coefficients and $X_d=X(t-T)$ is the delayed vector. 

We begin our analysis writing the above equation in one of the two representations, say the STR, and evaluating the system response to an external spatiotemporal perturbation $Y$,
\begin{equation}
X_\sigma = AX +BX(\sigma,\theta-1) +Y~.
\end{equation}
We then look for solutions in the Laplace domain for both variables after the transient related to the initial conditions. 

Denoting with $(s_\sigma,s_\theta)$ the Laplace-conjugate variables of $(\sigma,\theta)$ and with $\tilde{X}$ and $\tilde{Y}$ the transformed variables, we find
\begin{equation}
(s_\sigma I -A-e^{-s_\theta I} B )\tilde{X} = \tilde{Y}~,
\end{equation}
where $I$ is the identity matrix. We thus obtain the response of the system to the stimulus $Y$ in Laplace space
\begin{equation}
\tilde{X} = \chi(s_\sigma,s_\theta)~\tilde{Y}~,
\end{equation}
where we have defined the {\it susceptibility} matrix as
\begin{equation}
\chi(s_\sigma,\bar{\theta}) = (s_\sigma I -A-e^{-s_\theta I}B)^{-1}~.
\label{susc}
\end{equation}
Since the function (\ref{susc}) represents the system response to a unit impulse, it must satisfy Kramers-Kr\"{o}nig relations to obey causality (no response before the impulse is applied) \cite{toll}.

The Kramers-Kr\"{o}nig relations are valid for any function which is analytic in the upper-half complex plane and vanishes as $1/|s|$ or faster as $|s| \to \infty$, where $s$ is the Laplace-conjugate variable relative to the direction under consideration. One can readily verify that this is actually the case when considering the variable $\sigma$: indeed for $|s_\sigma| \to \infty$ and $s_\theta=const$, i.e. along the $\sigma$ direction, the susceptibility displays the asymptotic behavior
\begin{equation}
\chi \simeq  s_\sigma^{-1}I.
\end{equation}
On the other hand, along the $\theta$ direction, i.e. for $|s_\theta| \to \infty$ and $s_\sigma=const$, we find
\begin{equation}
\chi \simeq (s_\sigma I -A)^{-1}.
\label{chi2}
\end{equation}

A finite susceptibility (for each spatial frequency $s_\sigma $) at infinity along the $s_\theta$ axis has a precise physical meaning: the system equally responds at all temporal frequencies up to infinity. In the time domain, this would imply an unphysical instantaneous coupling (i.e. at the same $\sigma$ point) between a delay and the successive.

We can thus conclude that, in the susceptibility of the full problem (i.e. without any approximation) there exists a forbidden direction along the $\theta$ variable where the causality falls. As a consequence, one should consider the opportunity to use such a variable as equivalent to the physical time. In the next subsection, we support this interpretation by the analysis of the comoving Lyapunov exponent.

\subsection{Comoving Lyapunov Exponent}

The (maximum) comoving Lyapunov exponent (CLE) is an useful tool to characterize how a localized spatio-temporal disturbance propagates in different directions \cite{comoving}. In particular, it allows to determine how information is transmitted along lines in the domain of a pattern, shading light on their possible physical interpretation as causal routes.

In the following, we calculate it explicitly for the linear delay model 
\begin{equation}
\dot{z}(t) = -z(t) +\eta z(t-T)~,
\label{lineard}
\end{equation} 
using the method of chronotopic Lyapunov analysis  \cite{chrono}.

To this aim, we rewrite Eq. (\ref{lineard}) in the STR
\begin{equation}
\partial_\sigma Z(\sigma,\theta) = -Z(\sigma,\theta) +\eta Z(\sigma,\theta-1)~,
%\partial_\sigma Z(\sigma,\theta) = a Z(\sigma,\theta) +b Z(\sigma,\theta-1)~,
\label{lin-del-STR}
\end{equation} 
and in the DR as
\begin{equation}
\partial_\tau Y(\xi,\tau) = -Y(\xi,\tau) +\eta Y(\xi-1,\tau).
%\partial_\tau Y(\xi,\tau) = a Y(\xi,\tau) +b Y(\xi-1,\tau).
\label{lin-del-DR}
\end{equation} 

and we look for solutions of the type
\begin{equation}
Y(\xi,\tau)= Y_0\exp( {\bar{\mu} \xi + \bar{\lambda} \tau} )~,
\end{equation} 
where $\bar{\lambda} = \lambda +i \omega, \bar{\mu} = \mu +i \kappa$ (a similar ansatz can be used for the STR). Substituting the above solution into (\ref{lin-del-DR}) and separating real and imaginary part, we obtain
\begin{eqnarray}
\lambda     & = & -1 + \eta  e^{-\mu} \cos (\kappa) \\
\omega      & = & -\eta e^{-\mu} \sin (\kappa) ~.
\end{eqnarray}

The maximum LE for both the STR and DR is found at $\omega=0$ or, equivalently, $\kappa=0$. In the STR, the propagation velocity of the disturbance is
\begin{equation}
V_{STR} = -\frac{d\mu}{d\lambda} = \frac{1}{1+\lambda}~,
\end{equation}
and the corresponding CLE is
\begin{eqnarray}
\Lambda_{STR}(V_{STR}) &=& \mu+\lambda V_{STR} \nonumber \\
                       &=& 1-V_{STR}+\log(\eta V_{STR})~,
\end{eqnarray}
as reported in \cite{Giacomelli1996}.

In the case of the DR, the velocity is given by 
\begin{equation}
V_{DR} = -\frac{d\lambda}{d\mu} = \eta e^{-\mu}~,
\end{equation}
and the CLE by
\begin{eqnarray}
\Lambda_{DR}(V_{DR}) &=& \lambda+\mu V_{DR} \nonumber \\
                     &=& -1 +V_{DR}-V_{DR}\log(\frac{V_{DR}}{\eta})~.
\end{eqnarray}

\begin{figure}
\begin{center}
\includegraphics*[width=0.9\columnwidth]{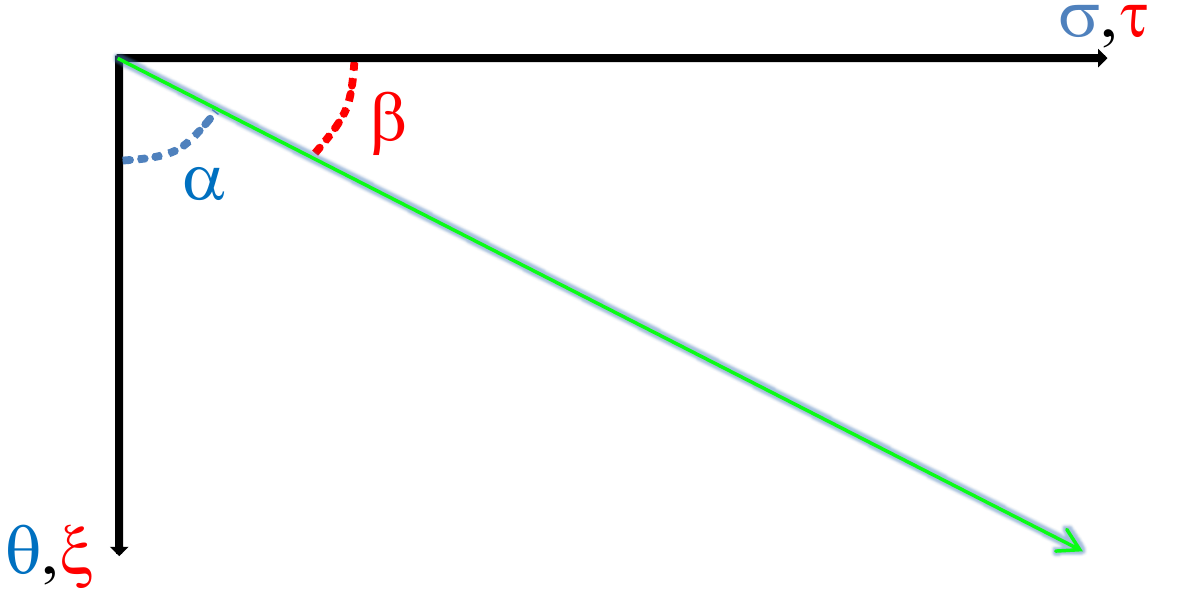}
\end{center} 
\caption{Geometric representation of the propagation angles in the STR and DR reference frame.}
\label{alfabeta}
\end{figure} 

The velocities in the two representations are thus related by 
\begin{equation}
V_{DR} =  \eta e^{-\mu} = 1+\lambda = \frac{1}{V_{STR}} ~,
\end{equation}
which can be interpreted geometrically (see Fig.\ref{alfabeta}) in terms of the relation between complementary propagation angles
\begin{equation}
V_{DR}=\tan \beta = \frac{1}{\tan \alpha}=\frac{1}{V_{STR}} ~.
\end{equation} 

The CLE are related as well by
\begin{equation}
\Lambda_{STR}(V_{STR}) = \frac{1}{V_{DR}}\Lambda_{DR}(V_{DR}) ~,
\end{equation}
or, equivalently
\begin{equation}
\Lambda_{DR}(V_{DR}) = \frac{1}{V_{STR}}\Lambda_{STR}(V_{STR}) ~.
\end{equation}

The above formulas relate the rates for a perturbation measured in the tangent space of the two representations for an arbitrary velocity, i.e. for a certain propagation direction of the perturbation. 

For instance, a spatio-temporal perturbation with characteristic width $\Delta \xi$ in the DR space-time ($\xi$,$\tau$) propagates in an interval $\Delta \tau = \Delta \xi / V_{DR}$. Writing
\begin{equation}
V_{STR} = \Delta \sigma /\Delta \theta = \Delta \tau /\Delta \xi =1/V_{DR}~,
\end{equation}  
we get
\begin{equation}
\Lambda_{STR}~\Delta \theta = \Lambda_{DR}~\Delta \tau ~.
\label{absspread}
\end{equation}
Eq. \ref{absspread} expresses the absolute spreading (or shrinking) of a perturbation as measured in the two representations along the vertical and horizontal directions, which results as an invariant. 

\begin{figure}
\begin{center}
\includegraphics*[width=1.\columnwidth]{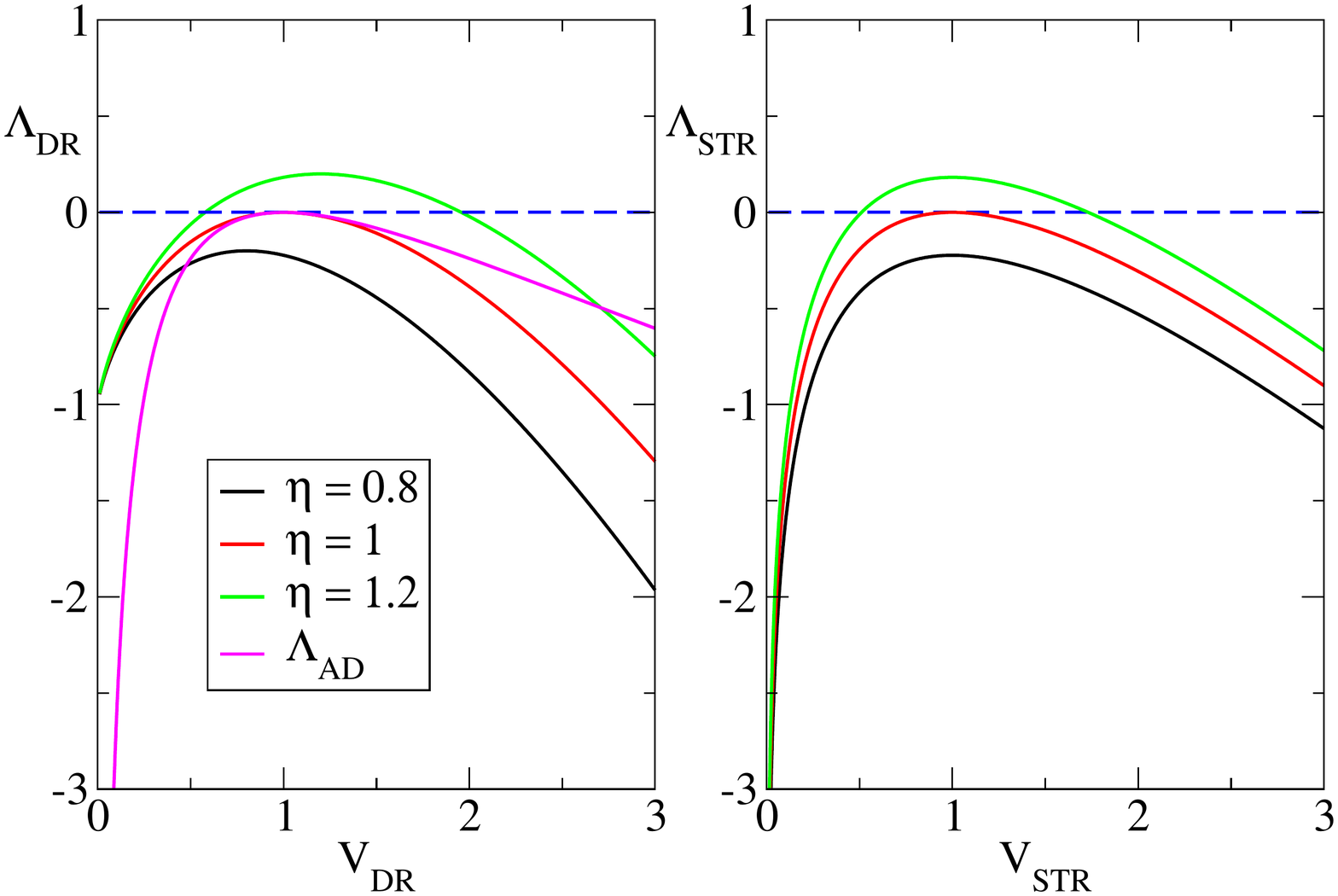}
\end{center} 
\caption{CLE for the DR and STR models in the stable ($\eta=0.8$), and unstable regimes ($\eta=1.2$). At the bifurcation $\eta=1$ it is also shown the exponent for the AD model as a function of $V_{DR}$ (see text).}
\label{comovLIN}
\end{figure} 

We plot the CLE for the DR and STR in Fig.\ref{comovLIN}, for different values of the feedback gain $\eta$ as a function of their velocities. The macroscopic observables represented by the correlation directions, defined by zeros of the CLE \cite{Giacomelli2000}, are the same. Indeed, if $\Lambda_{STR}(\bar{V}_{STR})=0$ for $V_{STR}=\bar{V}_{STR}$, then $\Lambda_{DR}(V_{DR}=1/\bar{V}_{STR})=0$ and viceversa. In both representations the negative velocities are not allowed, indicating the presence of a causality boundary. As already discussed in \cite{Giacomelli1996}, the STR curves displays a logarithmic divergence in zero, while the non-local coupling of the DR is mapped to infinity (thus removing the logarithmic divergence). In the STR, this in fact corresponds to instantaneous coupling between consecutive delays related the non-analytic (i.e. non-causal) susceptibility in the $\theta$ domain (\ref{chi2}).

As discussed in Sec. II a PDEM can be obtained expanding the non-local term up to a given order to approximate the delayed dynamics. We can thus evaluate the CLE for the various orders of a spatio-temporal approximation. 

We start defining $\psi(\mu) = e^{-\mu}$ and rewrite
\begin{equation}
\lambda_\psi  = -1 + \eta \psi(\mu)~,
\label{lambdapsi}
\end{equation}
leading to the velocity
\begin{equation}
V_{\psi} = -\frac{d\lambda_\psi}{d\mu} = -\eta \psi^\prime(\mu)~.
\label{vpsi}
\end{equation}

The CLE is then 
\begin{eqnarray}
\Lambda_\psi(V_\psi) &=& \lambda_\psi+\mu V_\psi \nonumber \\
                     &=& -1 + \eta \psi(\mu) -\mu V_\psi~.
\end{eqnarray}

One can therefore use the (\ref{lambdapsi}) and (\ref{vpsi}) to eliminate the auxiliary variables $\{\lambda_\psi,\mu\}$ to obtain eventually $\Lambda_\psi=\Lambda_\psi(V_\psi)$ for the chosen function $\psi$. In particular, we can treat in this way different orders of expansion of the non-local term. 

For a second-order expansion of the DR model, which corresponds to an advection-diffusion (AD) term, $\psi(\mu) = e^{-\mu} \approx 1-\mu+\frac{1}{2}\mu^2$. As a consequence, the velocity is given by
\begin{equation}
V_{AD} = -\frac{d\lambda}{d\mu} = \eta (1-\mu)~,
\end{equation}
and the CLE reads
\begin{eqnarray}
\Lambda_{AD}(V_{AD}) &=& \lambda+\mu V_{AD} \\
                     &=&   -1 +V_{AD}+\frac{\eta}{2}(\frac{V_{AD}}{\eta}-1)^2 \nonumber \\
                      &&    -V_{AD}(\frac{V_{AD}}{\eta}-1) \nonumber~.
\end{eqnarray}
In Fig. \ref{comovLIN} (left panel), we compare the above exponent at the bifurcation point $\eta=1$ with the CLE in the DR. The horizontal variable is evaluated by $V_{AD}=\eta(1+\log|{V_{DR}\over \eta}|)$. As seen from the plot, the advection-diffusion model is already a good approximation of the system around the maximum in terms of the CLE.

We finally remark that only the non-local or delayed term may induce problems with causality. Every finite-order PDE model is free from that, and it is increasingly correct at higher orders around the comoving direction (the location of maximum of the CLE).

\section{The delayed Adler equation}
\label{adler}

We now investigate and discuss the two representations in the framework of the so-called delayed Adler's equation. 
The model describes the evolution of the phase of the optical field in optically-injected laser systems with time-delay feedback and accounts for the formation and interaction of topological localized states \cite{garbin2015} (homoclinic $2 \pi$-kink solutions) very similar to those found in the Sine-Gordon equation. The model reads 
\begin{equation}
\dot{\phi} = \Delta -\sin\phi +\chi\sin(\phi_d-\phi -\psi) ~,
\label{dam}
\end{equation}
where $\phi$ is the phase of the optical field, $\Delta$ is proportional to the detuning between the injection and the laser frequency, $\chi$ is the normalized feedback strength and $\psi$ is related to the feedback phase.
For the purposes of this work, Eq. (\ref{dam}) just provide a non-trivial scalar system where the delayed feedback is nonlinear, thus leading to significant differences in the spatio-temporal representation with respect to models considered in \cite{Marino2018}. 

According to the DR, we obtain
\begin{equation}
\phi_\tau = \Delta -\sin\phi +\chi\sin \big(\phi_{NL}-\phi -\psi \big) ~,
\label{damDR}
\end{equation}

which together to suitable boundary conditions for $\phi$ along the $\xi$ domain, which we take periodic, represents the essence of our approach.

\begin{figure}
\begin{center}
\includegraphics*[width=1.\columnwidth]{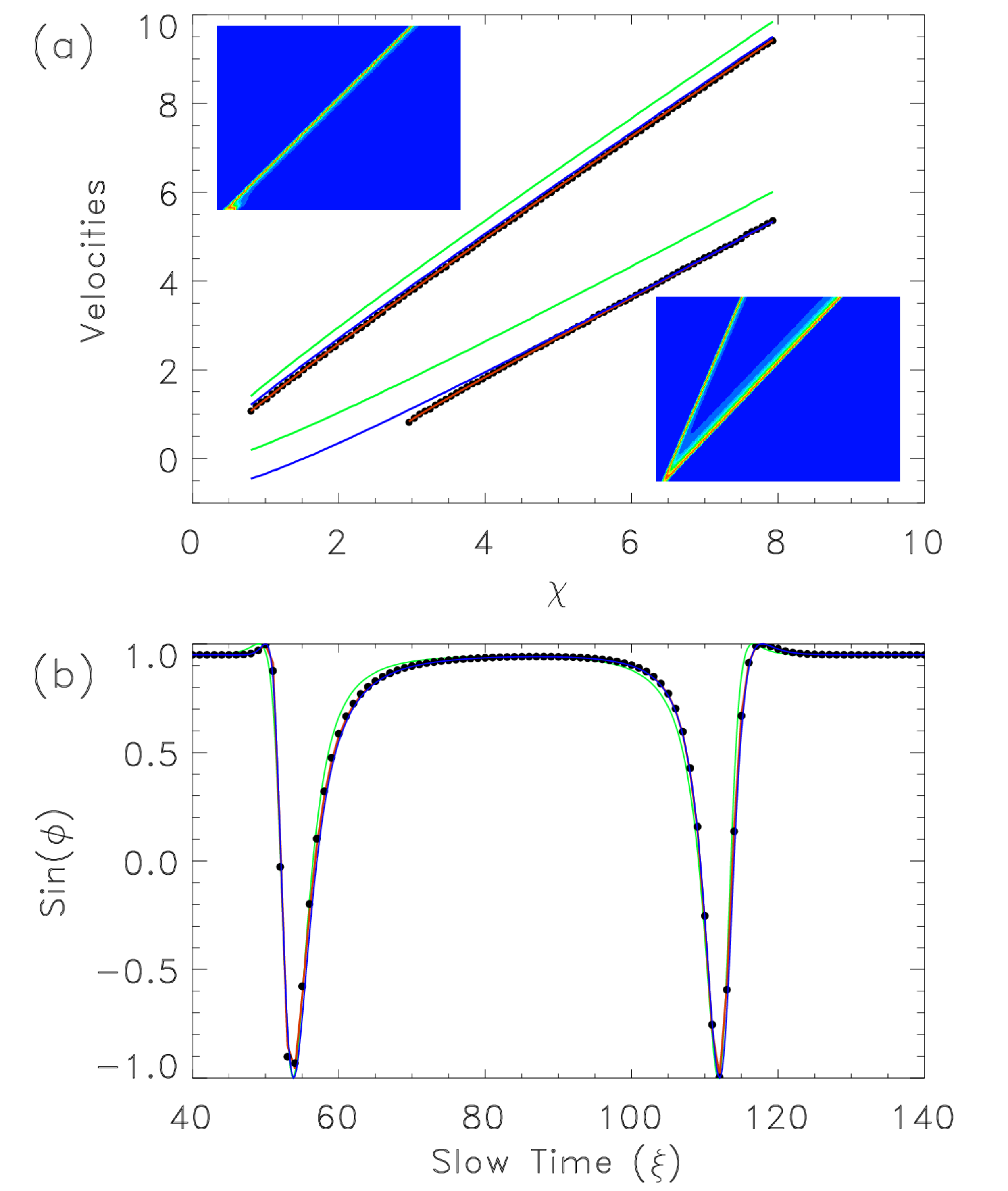}
\end{center} 
\caption{The delayed Adler model and its spatio-temporal descriptions. Top: Pulse propagation velocities in the $(\sigma,\theta)$ plane for the delayed (red) and the non-local (black dots) model together with its second-order (green) and third-order (blue) approximations in the DR. In the insets the 1-pulse (a) and two-pulses (b) solutions of the delayed Adler model are plotted in the DR for $\chi=2$ and $\chi=4$, respectively (here we plot the sinus of $\phi$). Bottom: Transverse cuts along the spatial $\xi$ direction for the different models: the color code is the same used for the top panel. Other parameters are $\Delta=0.95$, $T=2\times10^3$ and $\psi=0$.}
\label{cfr-vel}
\end{figure}

A normal form approximation of (\ref{damDR}) can be readily obtained by expanding the non-local term up to a given order. At the second-order we obtain
\begin{eqnarray}
\phi_\tau +\phi_\xi\chi\cos\psi&=& \Delta  -\chi\sin\psi -\sin\phi~  \nonumber \\
&&              +\frac{1}{2}\chi\sin\psi~\phi_\xi^2 +\frac{1}{2}\chi\cos\psi~\phi_{\xi\xi}~. \label{damDR2}
\end{eqnarray}
The comparison between the models (\ref{dam}), (\ref{damDR}) and (\ref{damDR2}) is reported in Fig.\ref{cfr-vel}. 
Starting from a rectangular initial condition, we identify two propagating regimes as the feedback strenght is varied: at low values of $\chi$, a single localized state propagating with constant velocity and, for higher values of the parameter, two pulses propagating at different speeds. The corresponding spatiotemporal patterns are shown in the insets of Fig.\ref{cfr-vel}(a) where we plot the sinus of the phase variable $\phi$. These coexixting localized states correspond to coarsening of kink-antikink solutions while the single pulse regime at low values of $\chi$ corresponds to the propagation of a single kink (phase-slip).

In \ref{cfr-vel}(a), we plot the velocities of the kinks for a decade range of the feedback gain parameter $\chi$. We observe an excellent agreement between the delayed (\ref{dam}) and the non-local model (\ref{damDR}), not only at the level of propagation speeds, but also in the transverse profiles of the solutions, as reported in Fig.\ref{cfr-vel}(b). Although this could appear somehow expected as the two models share the same bulk rule, we remark that they strongly differ at the boundaries. This supports our initial hypothesis that in the thermodynamic limit (i.e. in the bulk region) the non-local model well approximates the delayed dynamics, independetly from the choice of the boundary conditions.

The second-order normal form (\ref{damDR2}) captures most of the phenomenology of the delayed and non-local models, reproducing the qualitative behavior of the velocities as a function of $\chi$ (see Fig.\ref{cfr-vel}a) and also providing a good approximation of the profiles of the kinks (Fig.\ref{cfr-vel}b).
On the other hand, it does not display the transition to the single-kink regime at low values of $\chi$, and a noticeable difference in the magnitude of the velocities is observed. 

As a peculiar benefit of the DR, we can improve the quality of the normal form approximation by simply increasing the order of the expansion of the non-local term. We report in Fig.\ref{cfr-vel} the results obtained by integrating a third-order expansion normal form, obtained by adding to the r.h.s. of (\ref{damDR2}) the terms
\begin{equation}
\frac{1}{6}[(\chi\cos\psi~\phi_\xi^3 - \phi_{\xi\xi\xi}) - 3 \chi\sin\psi ~\phi_\xi\phi_{\xi\xi}]. 
\end{equation}
Even in this case, we the model is unable to reproduce the transition from the double to the single pulse regime. However, the agreement between both the velocities and the spatial profiles is substantially improved and barely distinguishable from those obtained from the full models Eqs. (\ref{dam}) and (\ref{damDR}). Here, the introduction of higher orders in the normal form breaks the parity symmetry of the solutions around the comoving direction, witnessed by the presence of only even terms in (\ref{damDR2}) besides the drift term, thus leading to a better approximation of the original, asymmetric profiles.

We finally observe that, writing Eq. (\ref{damDR2}) in the comoving reference frame corresponding to the velocity $v = \chi \cos \psi$ to remove the advection term in the l.h.s., and rescaling the space by 
\begin{equation}
\xi \to \xi\sqrt{\frac{1}{2}\chi \cos \psi}
\end{equation}

we eventually get
\begin{equation}
\phi_\tau=  \sin{\bar{\phi}}-\sin\phi +\phi_{\xi\xi} +\phi_\xi^2\tan\psi ~, 
\label{damNF}
\end{equation}

where $\sin{\bar{\phi}}= \Delta  -\chi\sin\psi$.  

The model (\ref{damNF}) is now formally identical to the second-order normal form equation obtained in Ref. \cite{garbin2015} in the STR [cf their Eq. (3)], and represents an alternative mathematical description of the system. However, in (\ref{damNF}) the role of time and space is exchanged: indeed for the advection velocity associated to the feedback term we find the value $\chi\cos\psi$ that is the inverse of what reported in \cite{garbin2015}. We will turn back to this issue in Sec. VI, where we will discuss the connection between the two representations based on general arguments.

\section{An optical delayed model with dispersion}
\label{disp}
In this section we deal with a rather interesting framework, formalized by a delayed differential equation with an algebric constraint, which is thus intermediate between a scalar and a vectorial case. The model has been first introduced in \cite{schelte2019} for the study of dispersive instabilities of pulse trains in mode-locked semiconductor lasers. Here, we specifically refer to the single-mode version in \cite{schelte2020} [see also our Eq. (\ref{fullmodel})], from which an equivalent PDE in the STR has been derived by means of multiple-scale analysis.

\subsection{The linear case}

Before discussing the full model (\ref{fullmodel}) we first examine a linear prototype system, in which already most of the topic can be elucidated,
\begin{eqnarray}
\dot{E} &=& -E +h Y \\
Y &=& \eta (E_{d}-Y_{d}) \nonumber~.
\end{eqnarray}
Here, we are interested in studying the above model in itself, regardless its physical meaning; notice however, that it could be obtained from (\ref{fullmodel}) by eliminating the carrier dynamics (i.e. neglecting all nonlinear terms).

Setting $u=E$ and $w=E-Y$ the model can be rewritten as
\begin{eqnarray}
\dot{u} &=& (h-1)u -h w \label{3disp}  \; \\
w         &=& u-\eta w_d \nonumber~. \\
\end{eqnarray}

In the framework of the DR, the above equation takes the form
\begin{eqnarray}
\partial_\tau{u} &=& (h-1)u -h w \label{3dispDR} \; \\
u         &=& u-\eta w_{NL} \nonumber~. \\
\end{eqnarray}

Fourier transforming we can derive the exact dispersion relation
\begin{equation}
p(q) = h-1-\frac{h}{1+\eta e^{-iq}}~,
\label{disp1}
\end{equation}
where we associate the frequency $p(q)$ and the wavevector $q$ to the time derivative and spatial shift operator, respectively, i.e. $\partial_\tau \to p(q)$ and $\textit{\bf S}_1 \to e^{-iq}$.

Expanding the exponential term for small wavevectors, e.g. up to the second order we get
\begin{eqnarray}
p(q) &=& \frac{(\eta-1)h-1}{\eta +1} -\frac{\eta h}{(\eta+1)^2}iq \\
      && -\frac{\eta(\eta-1)h}{2 (\eta+1)^3}(iq)^2 \nonumber~,
\end{eqnarray}
which corresponds in the direct spacetime ($\xi$,$\tau$) to the normal form
\begin{equation}
Z_\tau = \frac{(\eta-1)h-1}{\eta +1}Z -\frac{\eta h}{(\eta+1)^2}Z_\xi -\frac{\eta(\eta-1)h}{2 (\eta+1)^3}Z_{\xi\xi}~.
\end{equation}

Most of the interest for this model comes from the observation that for high reflectivities ($\eta \to 1^-$) the coefficient of the second-order spatial derivative, i.e. the diffusion, vanishes. In this limit, we are left with a dominant role of the dispersive effects related to the third order term. To study this regime, we set $h=2$ (corresponding to the Gires-Tournois interferometer regime \cite{schelte2019,gt}) and $\eta=1-\varepsilon$ with $\varepsilon =o(1)$, to obtain
\begin{equation}
p(q) = -\frac{1}{2}\varepsilon -\frac{1}{2}i q + \frac{1}{8}\varepsilon (i q)^2+\frac{1}{24}(i q)^3 +...
\label{disp2}
\end{equation}

Truncating the expansion at the third-order, the corresponding normal form writes as
\begin{equation}
U_\tau = -\frac{1}{2}\varepsilon U -\frac{1}{2} U_\xi +\frac{1}{8}\varepsilon U_{\xi^2}+\frac{1}{24} U_{\xi^3}~.
\label{disp3}
\end{equation}

\begin{figure}
\begin{center}
\includegraphics*[width=1.\columnwidth]{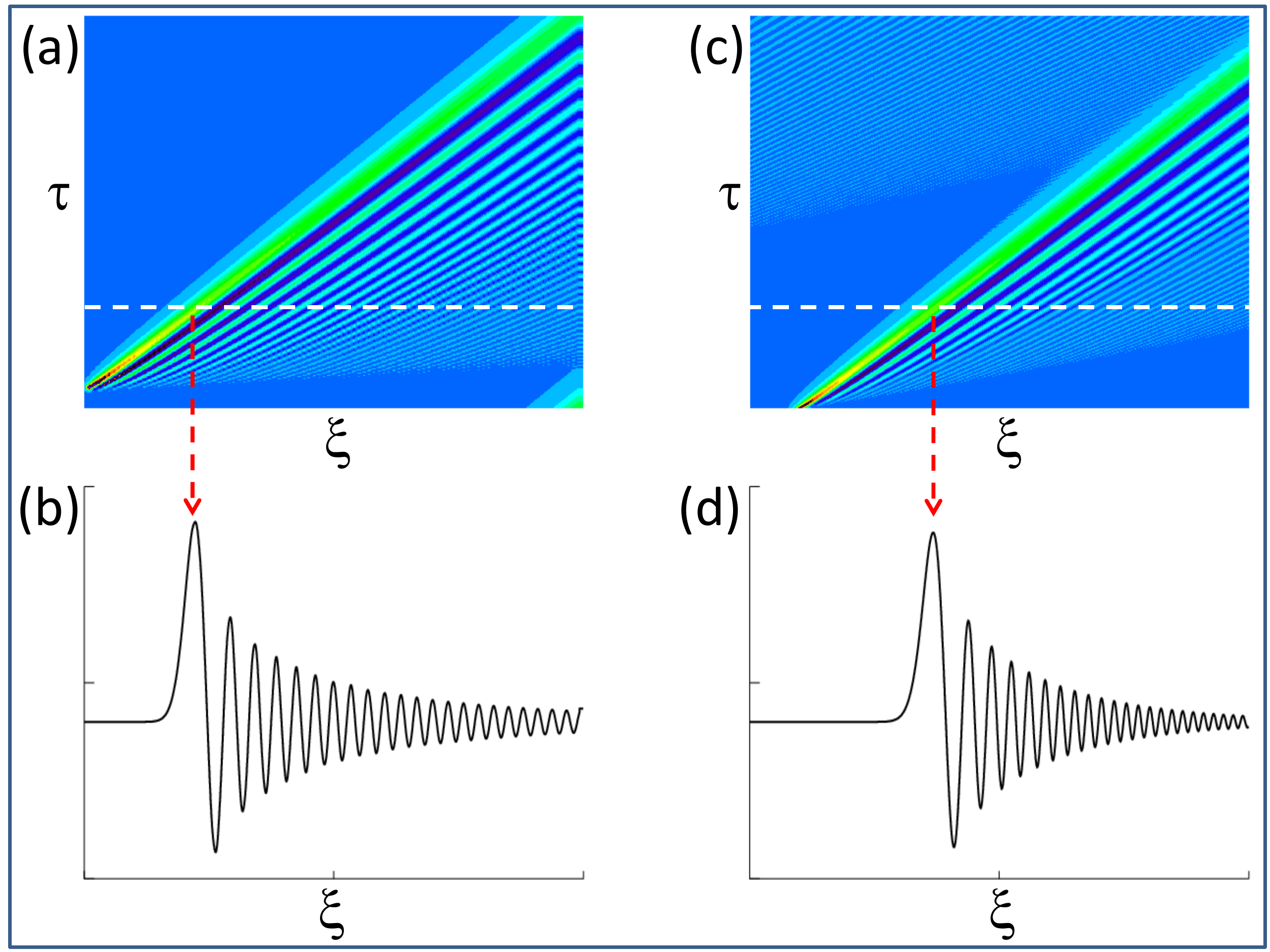}
\end{center} 
\caption{(a,b) Numerical integration of the delayed model (\ref{3disp}) and (c,d) of the spatially-extended system (\ref{disp4}) for $h=2$ and $T=200$. In (a,c) the spatiotemporal patterns are shown in the DR spacetime $(\xi,\tau)$. In (b,d) the profiles are transverse cuts along $\xi$ evaluated at fixed $\tau$, as indicated by the dashed lines). The vertical arrows indicate the pseudospace positions of the profiles maxima.}
\label{LD}
\end{figure}

In the limit $\varepsilon =0$ we get the dispersive-advection equation
\begin{equation}
U_\tau + \frac{1}{2}U_\xi = \frac{1}{24} U_{\xi^3}~.
\label{disp4}
\end{equation}
Eq. \ref{disp4} has the form of the linear Korteweg - De Vries (KdV) equation \cite{kdv}, although with a positive third-order coefficient. In optics, this corresponds to an anomalous dispersion term implying that the higher spatial-frequency waves travel faster than the lower frequency waves. The integration of model (\ref{disp4}) is in good agreement with the original delayed system (\ref{3disp}). The spatiotemporal plots also highlight the different boundary conditions of the two models. This is evidenced also looking at the maxima of the profiles that are found at different pseudospatial positions. Both the delay and the spatially extended model display anomalous dispersion effects with high-frequency components of the wavepacket propagating faster than the lower ones (see Fig. \ref{LD}). 

On the other hand, the KdV is often written with a negative third order coefficient, leading to the normal dispersion phenomena with lower frequency waves travelling faster. This is what we find in the STR description.
Reversing spatial and temporal variables
\begin{equation}
q(p) = -\log \Big( ~\frac{1+i p}{\eta( h-1-i p)} \Big)~,
\label{STRdisp1}
\end{equation}
and expanding for small $p$ up to the third-order we get the normal form for $h=2$
\begin{equation}
U_\theta = \log(\eta) U -2 U_\sigma -\frac{2}{3}U_{\sigma^3}~.
\label{STRdisp2}
\end{equation}
Eq. (\ref{STRdisp2}) corresponds to the normal form obtained in \cite{schelte2019} by means of functional mapping method \cite{schelteopt}.
By rescaling $\theta \to \theta/2$ and $\sigma \to 2\sigma$ and in the limit $\varepsilon =0$, we obtain
\begin{equation}
U_\theta = -\frac{1}{2} U_\sigma -\frac{1}{24}U_{\sigma^3}~,
\label{STRdisp3}
\end{equation}
which is the same PDEM found for the DR but with the opposite sign for the dispersion term. 

Remarkably, we have found that the {\it same} pattern can be generated with normal or anomalous dispersion when observed in the STR and in the DR, respectively. As a consequence, the very same bulk phenomena are expected to arise in the two PDEMs well approximating the original ones, but with opposite symmetry with respect to the diagonal axis. In the next section we will analyze these dispersive phenomena in the fully non-linear problem.   

\subsection{The nonlinear dispersive model}

We now consider the full model discussed in Ref. \cite{schelte2020}, which describes the dynamics of the intracavity field $E$ and of the population inversions $N_i$ ($i=1,2$) of a passively mode-locked integrated external-cavity surface-emitting laser
\begin{eqnarray}
\dot{E}   &=& \big( (1-i\alpha_1) N_1 +(1-i\alpha_2) N_2 -1\big)E + hY \label{fullmodel} \; \\ 
Y &=& \eta \big( E(t-T)-Y(t-T) \big) \nonumber\\
\dot{N_1} &=& \gamma_1(J_1 - N_1)- |E|^2N_1  \nonumber \\  
\dot{N_2} &=& \gamma_2(J_2 - N_2)- s|E|^2N_2 \nonumber \ ~. 
\end{eqnarray}
Here, $Y$ is the field in the external cavity, $\alpha_i$, $J_i$ and $\gamma_i$ are the linewidth enhancement factors, the bias and recovery time relative to the gain ($i=1$) and absorber section ($i=2$), respectively, and $s$ is ratio of the gain and absorber saturation intensities.

We rewrite (\ref{fullmodel}) in the DR, where, as usual, the delayed term becomes nonlocal in space and the standard time derivatives transform into derivatives with respect to the DR time $\tau$:
\begin{eqnarray}
\partial_\tau E &=& \big( (1-i\alpha_1) N_1 +(1-i\alpha_2) N_2-1\big)E + hY \label{FMnonloc} \; \\ 
Y               &=& \eta \big( E_{NL}-Y_{NL} \big) \nonumber\\
\partial_\tau N_1 &=& \gamma_1(J_1 - N_1)- |E|^2N_1   \nonumber\\  
\partial_\tau N_2 &=& \gamma_2(J_2 - N_2)- s|E|^2N_2 \nonumber ~. 
\end{eqnarray}

The $Y$ variable can be eliminated using the second equation in the Fourier domain
\begin{equation}
\bar{Y} = \bar{E} \frac{\eta e^{-iq}}{1 +\eta e^{-iq} }~, 
\end{equation}
where the bar indicates the Fourier transform and $q$ the spatial wavevector.

Expanding up to the third order and reverting to spatial variables we get
\begin{eqnarray}
Y &=& \big( ~\frac{\eta}{1+\eta} -\frac{\eta}{(1+\eta)^2}\partial_\xi 
            +\frac{\eta(1-\eta)}{2(1+\eta)^3}\partial_\xi^2             \\
   &&       +\frac{\eta(4\eta-1-\eta^2)}{6(1+\eta)^4}\partial_\xi^3  +.. ~\big)E~. \nonumber 
 \end{eqnarray}
 
Substituting into the full model, we obtain
\begin{eqnarray}
\partial_\tau E   &=& \big( (1-i\alpha_1) N_1 +(1-i\alpha_2) N_2\big)E  \label{FMthird} \; \\ 
                  && +(\frac{h\eta}{1+\eta} -1)E  \nonumber  \\
&& -\frac{h \eta}{(1+\eta)^2}\partial_\xi E+\frac{h\eta(1-\eta)}{2(1+\eta)^3}\partial_{\xi^2} E       \nonumber  \\
        && +\frac{h\eta(4\eta-1-\eta^2)}{6(1+\eta)^4}\partial_{\xi^3} E  \nonumber \\
\partial_\tau N_1 &=& \gamma_1(J_1 - N_1)- |E|^2N_1   \nonumber \\  
\partial_\tau N_2 &=& \gamma_2(J_2 - N_2)- s|E|^2N_2 \nonumber ~.
\end{eqnarray}

\begin{figure*}
\begin{center}
\includegraphics*[width=1.5\columnwidth]{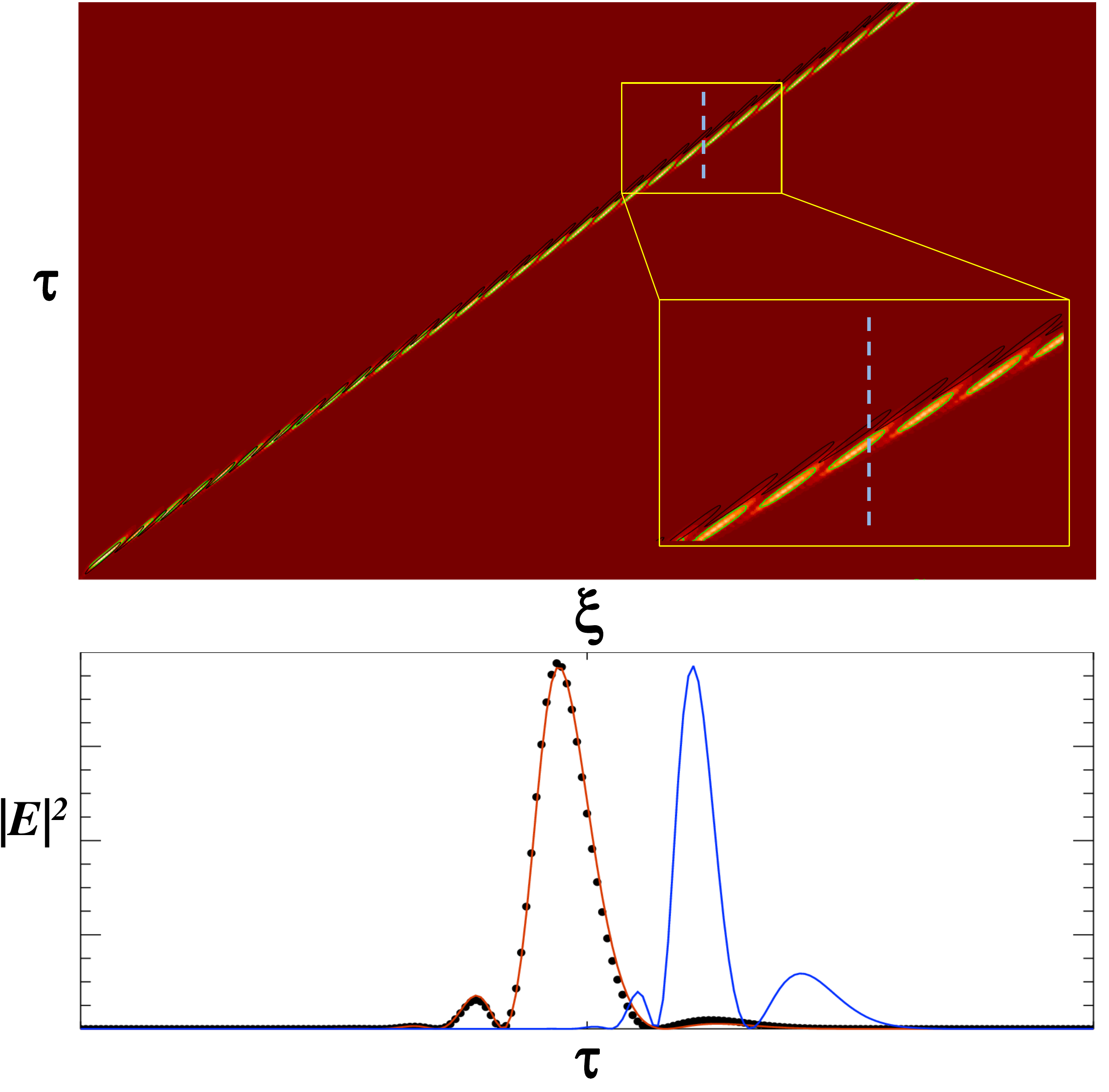}
\end{center} 
\caption{a) Spatio-temporal plot in the DR of for a single propagating pulse obtained by numerical integration of the full delay model (\ref{fullmodel}) (yellow pattern), the nonlocal DR model (\ref{FMnonloc}) (black solid line) and its third-order approximation (\ref{FMthird}) (green solid line). b) Transverse cuts of the above patterns at fixed $\xi$ indicated by the vertical dashed lines.  
Parameters: $\alpha_{1,2}$=0, $J_1=0.65$, $J_2=-0.5$ and $\gamma_1=3\times10^{-3}$, $\gamma_2=0.1$, $\eta=0.7$, $h=2$ and $s=1$. In the delay model $T=2000$.}
\label{nonlin-disp}
\end{figure*} 

In Fig. \ref{nonlin-disp} we compare the results obtained by numerical integration of the delay system (\ref{fullmodel}), the corresponding nonlocal DR model (\ref{FMnonloc}) and the third-order PDEM (\ref{FMthird}). The spatiotemporal plot shows the propagation of a single pulse in the DR, the dynamics of which have been analyzed in detail in Ref. \cite{schelte2020}. The agreement between the delayed and DR model is excellent. On the other hand, while well reproducing the phenomenology, the third-order approximation exhibits quantitive differences, both in the propagation velocity and in the transverse profiles (see Fig. \ref{nonlin-disp}b).

We now compare our PDEM with the results of \cite{schelte2020}, where a third-order model in the STR has been derived. 
Rescaling the pseudo-space $\xi$ by ${ (1+\eta)^2 / h \eta}$ and the pseudo-time $\tau$ by $h \eta/(1+\eta)^2$, we have
\begin{eqnarray}
\partial_{\xi^n} \to \frac{(1+\eta)^{2n}}{(h \eta)^n} \partial_{\xi^n}~,~n=1,2,3 \\
\partial_{\tau} \to  h \eta/(1+\eta)^2 \partial_{\tau}~,
\end{eqnarray}
and we eventually obtain
\begin{eqnarray}
\partial_\tau E   &=& \big( (1-i\alpha_1) N_1 +(1-i\alpha_2) N_2  \label{FMthird2} \; \\
                  && +\frac{h\eta}{1+\eta} -1\big)\frac{h \eta}{(1+\eta)^2}E  \nonumber \\
&& -\frac{(1+\eta)^{2}}{h \eta}\partial_\xi E \nonumber \\
&& +\frac{(1-\eta^2)}{2} \big(\frac{1+\eta}{h \eta} \big)^2 \partial_{\xi^2} E       \nonumber \\
&& +\frac{4\eta-1-\eta^2}{6(1+\eta)^2}\frac{(1+\eta)^{6}}{(h \eta)^3} \partial_{\xi^3}E \nonumber \\
\partial_\tau N_1 &=& \gamma_1(J_1 - N_1)- |E|^2N_1  \nonumber \\  
\partial_\tau N_2 &=& \gamma_2(J_2 - N_2)- s|E|^2N_2 \nonumber ~. 
\end{eqnarray}
It is interesting to note that apart from the usual exchange between space and time, the drift (first-order) and diffusion (second-order) terms are equal to those found in the STR model in \cite{schelte2020} (see their Eq. (11) for the model and Eqs. (14-15) for the drift and diffusion coefficients, respectively). On the other hand, the coefficient of the third-order derivative is

\begin{eqnarray}
d_{DR}=\frac{4\eta-1-\eta^2}{6(1+\eta)^2}\frac{(1+\eta)^{6}}{(h \eta)^3} 
&=&   \\\frac{(4\eta-1-\eta^2)(1+\eta)}{2(1+\eta^3)} 
^^ \times \frac{(1+\eta^3)}{3} \big(\frac{1+\eta}{h \eta}\big)^3 &=& \nonumber\\
 Q(\eta) \times \frac{(1+\eta^3)}{3} \big(\frac{1+\eta}{h \eta}\big)^3 &=& -Q(\eta)d_3 \nonumber ~,
\end{eqnarray}
where $d_3$ is the dispersion constant in \cite{schelte2020} [cf their Eq. (16)]. The two dispersions thus differ both in sign and absolute value. Interestingly however, since $\eta =1-\varepsilon$ in the high reflectivity limit $\varepsilon \ll 1$ for which the diffusion vanishes we have
\begin{equation}
Q(\eta) \approx 1 - \frac{3}{2} \varepsilon^2
\end{equation}
Hence, up to second order corrections, or for the ideal case of perfect reflectivity $\varepsilon=0$, the two coefficients will only differ in sign as found in the case of the linear model.
We also notice that when $\eta=2-\sqrt{3} \approx 0.268$, $d_{DR}=0$ while $d_3$ is finite. As such, for this value of the reflectivity, the DR model (\ref{FMthird2}) is dispersionless while the STR model in \cite{schelte2020} remains dispersive. We remark that at the nonlocal level, including i.e. all infinite orders of the expansion, both the STR and DR models must coincide and reproduce the delayed dynamics in the thermodynamic limit. On the other hand, the rate at which the two representations converge towards the solution of the nonlocal problem is generally different and depends on the specific details of the system under consideration. In this case and for this value of reflectivity, higher-order derivatives are necessary for the DR model to capture dispersive effects. 

\section{Swapping space and time: from STR to DR and back}
\label{parity}

As we have seen in the previous sections, an effective approach to describe a long-delayed systems is to derive PDE's from the two representations. The task is to explicitly rule the evolution of the field variable along the temporal-like direction in terms of the space-like derivatives of it. Such a scheme can be very convenient, both from a conceptual and practical point of view. Indeed, whenever the model is obtained by means a suitable expansion, a few terms could be sufficient to well approximate the dynamics.

Since the role of pseudo-space and time is exchanged in the two representations, the function expressing the time derivative in terms of the space derivatives can be different in the two cases. As a consequence, the related PDEMs would differ as well, at least from a certain order on. 

In the following we will consider a specific class of PDEMs allowing to easily switch between the representations. In particular, we will find the conditions in which is possible to obtain the same reduced PDEM, obviously limited to some order in the space derivatives.

\begin{figure}
\begin{center}
\includegraphics*[width=0.9\columnwidth]{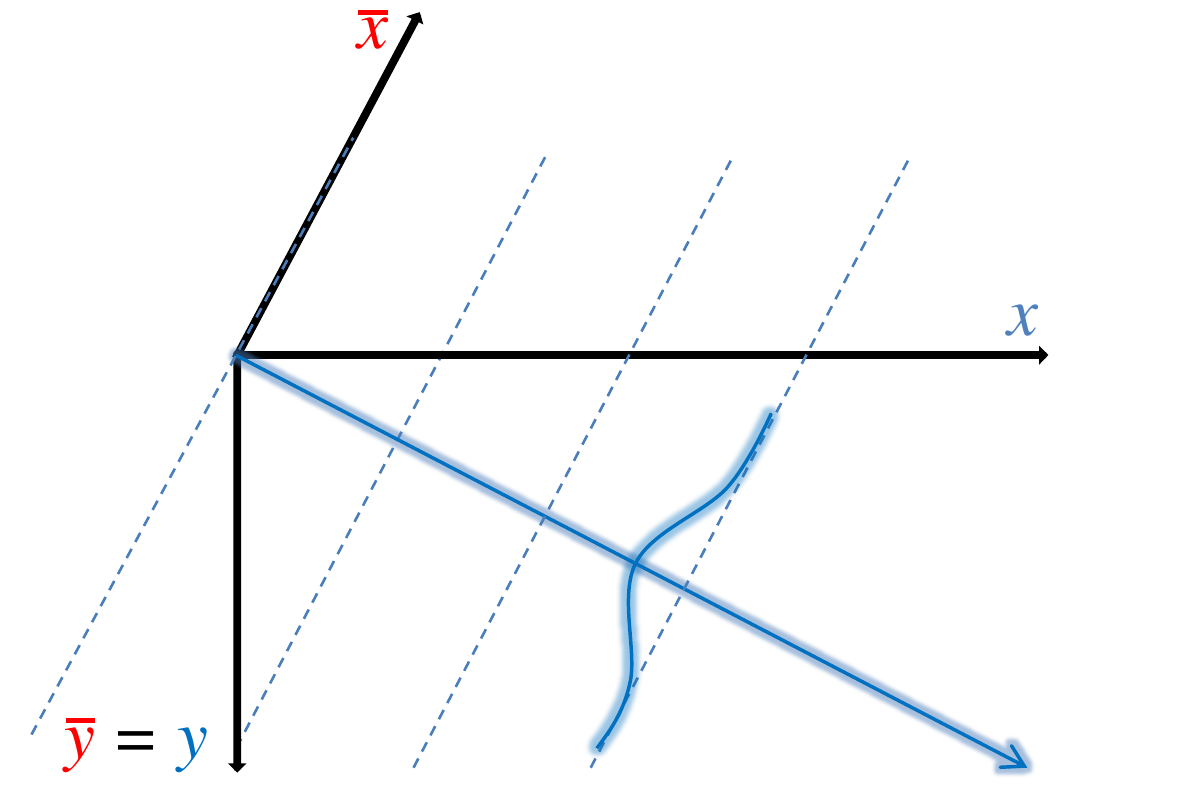}
\end{center} 
\caption{Representation of the original and comoving reference systems related by transformation (\ref{comfrc}).}
\label{symm}
\end{figure} 

We start from the $n$-order description obtained from the delay system in one of the two representations

\begin{equation}
\Phi_{y}^{(n)} = F^{(n)}(\Phi^{(n)},\Phi_{x}^{(n)},\Phi_{x^2}^{(n)},..,\Phi_{x^n}^{(n)})~.
\label{pdeSTR}
\end{equation}

We now move to the comoving reference frame corresponding to the diagonal of the 2D domain of the (suitably rescaled) variables. A pictorial view of these two reference frames is illustrated in Fig.~\ref{symm}.
\begin{eqnarray}
\bar{x} &=& x-y  \label{comfrc}  \\
\bar{y} &=& y   ~. \nonumber \\
\nonumber
\end{eqnarray}

Accordingly, Eq.~(\ref{pdeSTR}) rewrites as
\begin{equation}
\Phi_{\bar{y}}^{(n)} = \Phi_{\bar{x}}^{(n)}+F^{(n)}(\Phi^{(n)},\Phi_{\bar{x}}^{(n)},\Phi_{\bar{x}^2}^{(n)},..,\Phi_{\bar{x}^n}^{(n)})~.
\label{c-STR}
\end{equation}

The parity symmetry transformation $\bar{x} \leftrightarrow -\bar{x}$ which leaves invariant the above equation corresponds to {\bf $x \rightarrow y$} as seen by the comoving reference frame (\ref{comfrc}) and represents the commuting rule between the two representations. 

We now consider the class of model for which the comoving term, expressed by the linear first-order derivative, disappears from the equation: in the very common case of $n=2$ the equation (\ref{c-STR}) is thus invariant under (\ref{comfrc}). Therefore the transformed equation, which corresponds to the PDEM in the other representation is formally the same and admits the same solutions. In this case the parity transformation maps one representation, and its related solutions, into the other.

As a paradigmatic example in this class of second-order systems, we mention the Delayed Complex Landau (DCL) model
\begin{equation}
\dot{u} = \mu u -(1+i \beta)|u|^2 u +\eta u_d~,
\label{dcl}
\end{equation} 

where $u$ is complex. Once re-written in the DR and expanded up to the second-order yields
\begin{equation}
Z_\tau = (\mu +\eta) Z -\eta Z_\xi +\frac{\eta}{2} Z_{\xi\xi} -(1+i \beta)|Z|^2 Z~,
\label{dcl-DR-exp}
\end{equation}

i.e. a Complex Ginzburg-Landau (CGL) equation with drift $\eta$ and diffusion ${\eta /2}$. 

The corresponding second-order normal form for (\ref{dcl}) in the STR was obtained in \cite{Giacomelli1996} and can be written as
\begin{equation}
\eta Z_\theta = \mu_1 Z -Z_\sigma +\frac{1}{2 \eta} Z_{\sigma\sigma} -(1+i \beta)|Z|^2 Z~,
\label{dcl-STR}
\end{equation}
(see their Eq.(17) when reported in the original coordinate system).
Eqs.(\ref{dcl-DR-exp}) and (\ref{dcl-STR}) are identical setting  $\mu+\eta = \mu_1$ and 
\begin{eqnarray}
\partial_\tau &\to& \eta \partial_\theta  \label{swap-dcl} \;  \\ 
\partial_\xi  &\to& \frac{1}{\eta} \partial_\sigma ~.\nonumber \\
\nonumber
\end{eqnarray}

As seen, Eq.(\ref{swap-dcl}) corresponds to swap space and time between the two models and thus passing from STR to DR, with the proper units assured by the presence of the "velocity" $1/\eta$. 

We remark how, as long as the normal forms  (\ref{dcl-DR-exp})  and (\ref{dcl-STR}) derived from the two representations can be exchanged making use of the (\ref{swap-dcl}), the patterns obtained from the integration of the two are expected to be analogous and close to the one produced by the DCL model. This is true close to the Hopf bifurcation, but moving away from that this is no more the case as shown in \cite{Marino2018}. Indeed, the inclusion of a further term in the expansion (\ref{dcl-DR-exp}) allowed to better approximate the original dynamics including deviations from the parity symmetry observed moving away from the Hopf bifurcation. 

The bistable system with delay \cite{Giacomelli2012,Giacomelli2013}, with e.g. a quartic potential with asymmetry $a$ 
\begin{equation}
\dot{u} = -u(u-1)(u+1+a) +gu_d = -U^\prime(u) +gu_d
\label{ourbist}
\end{equation}
also belongs to this class of models. Again, it is straightforward to write the expansion in the DR at any order. At the second-order, we have the reaction-diffusion system
\begin{equation}
Z_\tau = -U^\prime(Z) +g Z - gZ_\xi + \frac{1}{2}gZ_{\xi \xi }~,
\label{2ndDR}
\end{equation}
and thus we set a correspondent STR model by using the rescaling (\ref{swap-dcl}) adopted for the DCL (with $\eta=g$)
\begin{equation}
gZ_\theta = -U^\prime(Z) +g Z - Z_\sigma + \frac{1}{2g}Z_{\sigma \sigma }~.
\label{2ndSTR}
\end{equation}

At the same order, the simplest nonlinear case occurs when the first order derivatives appear at second order power. This is indeed what we have found for the Adler model discussed in Sec. \ref{adler} where a term $\Phi_x^2$ is present and again the same PDEM is obtained in the two representations.

More complicate situations can arise for higher orders. For e.g. $n=3$, we can associate different functions for the two representation only differing by the sign of the third order derivative:
\begin{equation}
\Phi_{x}^{(3)} = F^{(3)}(\Phi^{(3)},\Phi_{x}^{(3)},\Phi_{x^2}^{(3)},\Phi_{x^3}^{(3)})~,
\end{equation}
for the first representation and 
\begin{eqnarray}
&& \Phi_{y}^{(3)} = F^{(3)}(\Phi^{(3)},\Phi_{y}^{(3)},\Phi_{y^2}^{(3)},-\Phi_{y^3}^{(3)}) \\
&& := \bar{F}^{(3)}(\Phi^{(3)},\Phi_{y}^{(3)},\Phi_{y^2}^{(3)},\Phi_{y^3}^{(3)})
\end{eqnarray}
for the second. In this way, the switch $x \to y$ has to be accompanied by $F \to \bar{F}$. This is what we have found in the Sec. \ref{disp}, where the two PDEMs only differ for the third order coefficient sign.

We finally remark that the PDEMs for the two representations are generally different also in the very simple cases. For instance, we consider again the linear equation (\ref{lineard}). Once written according to the DR in the Laplace-domain we obtain
\begin{equation}
s_\tau = -1 +\exp(-s_{\xi}) ~,
\end{equation}
where $(s_\xi,s_\tau)$ the Laplace-conjugate variables of $(\xi,\tau)$. Expanding up to the third order we have
\begin{equation}
s_\tau \approx -s_\xi+\frac{1}{2} s_{\xi}^2 -\frac{1}{6} s_{\xi}^3~, 
\end{equation}
which leads to the normal form
\begin{equation}
\Phi_\tau = -\Phi_{\xi} +\frac{1}{2} \Phi_{\xi\xi} -\frac{1}{6} \Phi_{\xi\xi\xi}~. 
\label{DRlin2}
\end{equation}

In the STR, we obtain instead for the corresponding conjugate variables
\begin{equation}
s_\sigma = -1 +\exp(-s_{\theta})~,
\end{equation}
and expanding
\begin{equation}
s_{\theta} = -\log(1+s_\sigma) \approx -s_{\sigma}+\frac{1}{2} s_{\sigma}^2 -\frac{1}{3} s_{\sigma}^3 ~, \end{equation}

we eventually get the normal form
\begin{equation}
\Phi_\theta = -\Phi_{\sigma} +\frac{1}{2} \Phi_{\sigma\sigma} -\frac{1}{3} \Phi_{\sigma\sigma\sigma}~.
\label{STRlin2} 
\end{equation}

\begin{figure}
\begin{center}
\includegraphics*[width=1.\columnwidth]{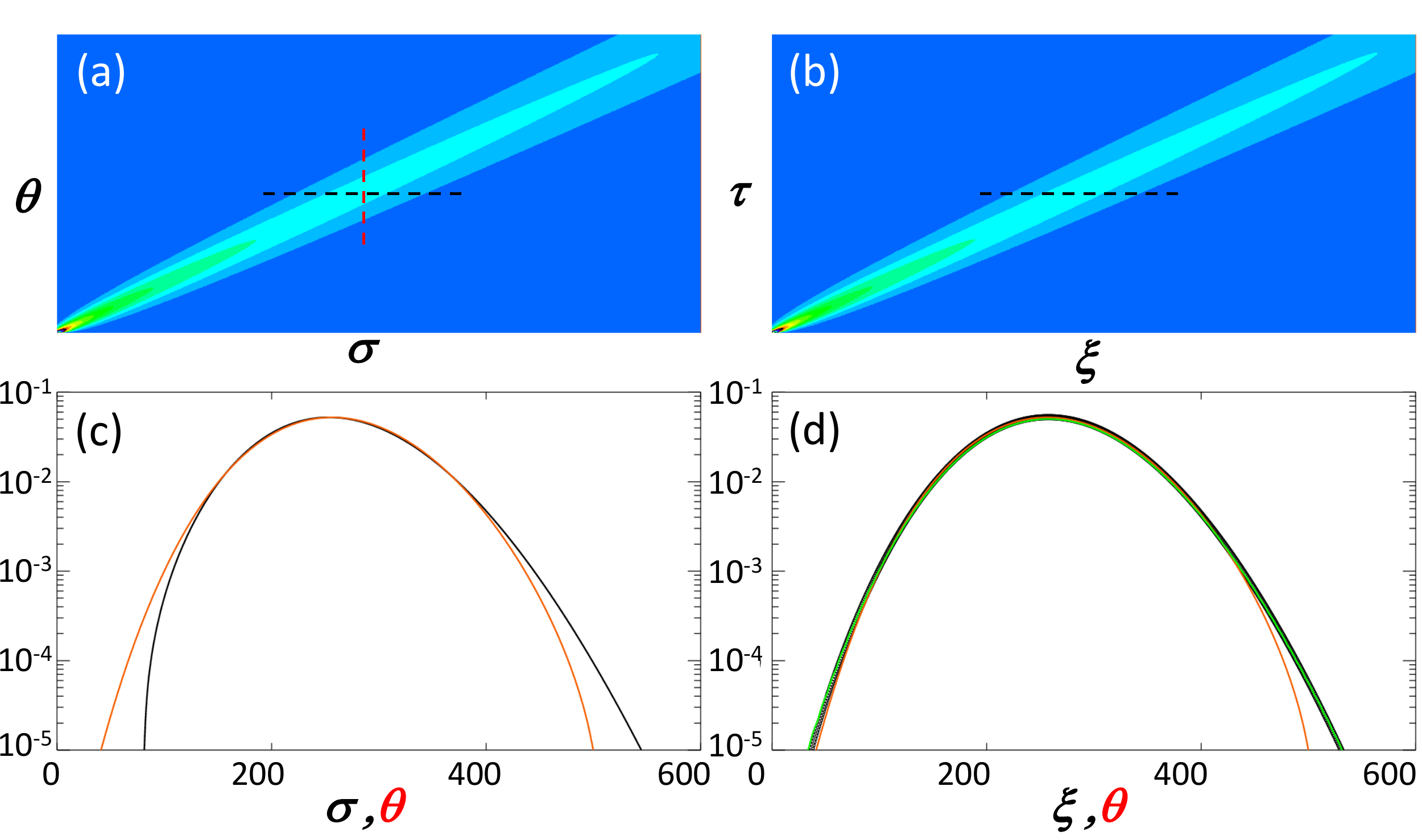}
\end{center} 
\caption{Spatiotemporal plots of the linear models (a) (\ref{STRlin2}) and (b) (\ref{DRlin2}) for a gaussian initial condition of unitary amplitude and width. (c) Profiles along the $\sigma$ (black line) and $\theta$ directions (red line) as obtained from model (\ref{STRlin2}). (d) Comparison between the profile along $\theta$ as obtained from (\ref{STRlin2}) (red line), the profile along $\xi$ as obtained from (\ref{DRlin2}) (black dots) and the solution of the delay model (\ref{lineard}) (green line). The profiles are displayed for the same value of $\sigma \leftrightarrow \tau$. The spatially extended models have been integrated using periodic boundary conditions.}
\label{lineardisp}
\end{figure}

We report in Fig. \ref{lineardisp} the numerical integration of the two models (\ref{DRlin2}) and (\ref{STRlin2}), in both cases using periodic boundary conditions and a gaussian initial function. As seen in Fig. \ref{lineardisp}a-b, the spatiotemporal patterns plotted in their respective spatiotemporal domains are quite similar. 
In Fig. \ref{lineardisp}c we compare the temporal and spatial profiles of system (\ref{STRlin2}). Due to dispersion, the two profiles are clearly asymmetric and different one from each other. A similar situation is found in model (\ref{DRlin2}), although with a weaker asymmetry owing to the lower dispersion coefficient. On the other hand, a remarkable agreement, up to almost three decades, is observed when we compare the two models along the corresponding directions, ($\theta \rightarrow \xi$) and ($\sigma \rightarrow \tau$), thus demonstrating the equivalence of the two representations. As an example, we plot in Fig. \ref{lineardisp}d the profile along $\theta$ shown in Fig. \ref{lineardisp}c, and a transverse cut along $\xi$ of the pattern in Fig. \ref{lineardisp}b. The residual deviations can be associated to the order of the expansion used in the two PDEs. Indeed, the two representations do not uniformly converge towards the solution of the delayed model and the inclusion of suitably different orders would be necessary to compensate the discrepancy in the profiles. In Fig. \ref{lineardisp}d we also plot the solution of the delayed system (\ref{lineard}) for comparison. Incidentally, in this specific case the DR model (\ref{DRlin2}) shows already at the third-order an excellent agreement with the original delay equation. However, deviations are eventually expected, as any finite-order expansion cannot capture the intrinsically nonlocal nature of the delay problem.  

\section{Conclusions}
The study of long delayed dynamical systems strongly benefits of a spatio-temporal description whenever it is possible. Such a mapping, besides realizing a bridge between different high-dimensional systems allows for simple conceptual interpretation of complicated phenomena otherwise hidden in the temporal series of a delayed system. As such, the success of the now widespread STR is explained and justified. However, some practical difficulties arise in the derivation of normal forms in the STR, as the implicit non-locality in time leads to involved mathematical derivations often requiring vicinity to a bifurcation. Moreover, both the evaluation of the comoving Lyapunov exponent and analytical considerations in linear models indicates that the choice of the slow-time variable as the pseudo-time in the representation could not be the most appropriate. In the spirit to better understand and describe the spatio-temporal equivalence, recently the new DR has been introduced. According to this new approach, the role of space and time is reversed in the mapping, aiming to describe a far from boundaries (bulk) evolution. While mathematically speaking the DR admits the very same solution of the original delay problem only for a specific choice of the boundary conditions, in the thermodynamic limit it is shown that the DR provides a very good approximation of the dynamics.

In this work, we have supported the preliminary arguments and evidences of the validity of DR over the STR by the analysis of new systems (with a nontrivial structure of the delayed feedback) and discussed in details the novelties and the peculiarities of this new approach. In particular, the easy derivation of PDEM at any order allows for a straightforward application of the method to describe the bulk dynamics of any long delayed systems. 

 We believe that with the new examples and enlightenment carried by this work our approach could represent a significant advance in the area of long-delayed dynamical systems. In particular, one can expect this to happen in the relevant cases of conceptual description and quantitative evaluation of bulk behaviours and quantities analogous to those of spatio-temporal systems.

Further investigations remain to be carried out, to precise the limit of application and the a-priori degree of approximation one could expect for a specific expansion.

\end{document}